\documentclass[preprint,12pt]{aastex}

\shortauthors{Winn et al.~2007}
\shorttitle{Prograde Orbit of TrES-2b}

\begin{document}

%
\def\ltsima{$\; \buildrel < \over \sim \;$}
\def\lsim{\lower.5ex\hbox{\ltsima}}
\def\gtsima{$\; \buildrel > \over \sim \;$}
\def\gsim{\lower.5ex\hbox{\gtsima}}
                                                                                          
%

\bibliographystyle{apj}

\title{
The Prograde Orbit of Exoplanet TrES-2b\altaffilmark{1}
}

\author{
Joshua N.\ Winn\altaffilmark{2},
John Asher Johnson\altaffilmark{3,4},
Norio Narita\altaffilmark{5},
Yasushi Suto\altaffilmark{5},\\
Edwin L.\ Turner\altaffilmark{6},
Debra A.\ Fischer,\altaffilmark{7},
R.\ Paul Butler\altaffilmark{8},
Steven S.\ Vogt\altaffilmark{9},\\
Francis T.\ O'Donovan\altaffilmark{10},
B.\ Scott Gaudi\altaffilmark{11}
}

\altaffiltext{1}{Data presented herein were obtained at the W.~M.~Keck
  Observatory, which is operated as a scientific partnership among the
  California Institute of Technology, the University of California,
  and the National Aeronautics and Space Administration, and was made
  possible by the generous financial support of the W.~M.~Keck
  Foundation.}

\altaffiltext{2}{Department of Physics, and Kavli Institute for
  Astrophysics and Space Research, Massachusetts Institute of
  Technology, Cambridge, MA 02139}

\altaffiltext{3}{Department of Astronomy, University of California,
  Mail Code 3411, Berkeley, CA 94720}

\altaffiltext{4}{Present address: Institute for Astronomy, University
  of Hawaii, 2226 Woodlawn Drive, Honolulu, HI 96822}

\altaffiltext{5}{Department of Physics, The University of Tokyo, Tokyo
  113-0033, Japan}

\altaffiltext{6}{Princeton University Observatory, Peyton Hall,
  Princeton, NJ 08544}

\altaffiltext{7}{Department of Physics and Astronomy, San Francisco
  State University, San Francisco, CA 94132}

\altaffiltext{8}{Department of Terrestrial Magnetism, Carnegie
  Institution of Washington, 5241 Broad Branch Road NW, Washington
  D.C.\ 20015-1305}

\altaffiltext{9}{UCO/Lick Observatory, University of California at
  Santa Cruz, Santa Cruz, CA 95064}

\altaffiltext{10}{California Institute of Technology, 1200 E.\
  California Blvd., Pasadena, CA 91125}

\altaffiltext{11}{Department of Astronomy, Ohio State University, 140
  W.~18th Ave., Columbus, OH 43210}

\begin{abstract}
  
  We monitored the Doppler shift of the G0V star TrES-2 throughout a
  transit of its giant planet. The anomalous Doppler shift due to
  stellar rotation (the Rossiter-McLaughlin effect) is discernible in
  the data, with a signal-to-noise ratio of 2.9, even though the star
  is a slow rotator. By modeling this effect we find that the planet's
  trajectory across the face of the star is tilted by $-9 \pm 12$~deg
  relative to the projected stellar equator. With 98\% confidence, the
  orbit is prograde.

\end{abstract}


\section{Introduction}

A small fraction of Sun-like stars have giant planets with orbital
periods smaller than about 10 days (Marcy et al.~2005, Udry \& Santos
2007). The existence of these planets was a surprise, because it was
expected that giant planets would only be found beyond the ``snow
line,'' with orbital distances greater than a few astronomical
units. Other surprises have come from detailed studies of individual
objects. Some are found on highly eccentric orbits (Johnson et
al.~2006, Bakos et al.~2007, Maness et al.~2007, Johns-Krull et
al.~2008). Some have mean densities that are quite small (Knutson et
al.~2007, Mandushev et al.~2007) or large (Sato et al.~2005, Torres et
al.~2007) in comparison with Jupiter.

However, in at least one sense, the close-in giant planets have
fulfilled prior expectations: they orbit their host stars in the
prograde direction, relative to the sense of the stellar
rotation. This is true, at least, of the 6 systems for which
measurements of spin-orbit alignment have been reported (Queloz et
al.~2000; Wolf et al.~2007; Narita et al.~2007a,b; Loeillet et
al.~2007; Winn et al.~2005, 2006, 2007a). In all of these cases but
one, the sky projections of the orbital axis and the stellar rotation
axis are observed to be fairly well-aligned, with measurement
precisions ranging from about 1.5 to 30 deg. The exception is
HD~17156, for which the angle between those axes was found to be
$62\pm 25$~deg (Narita et al.~2007b). In all of these cases, the
measurement technique relies upon the Rossiter-McLaughlin (RM) effect,
the anomalous Doppler shift that occurs during transits due to stellar
rotation (see, e.g., Queloz et al.~2000, Ohta et al.~2005, Gim\'enez
2006, Gaudi \& Winn 2007, Winn 2007).

A close alignment between the orbital and rotational axes seems
natural because this pattern prevails in the Solar system, and because
the angular momenta of the parent star and the planetary orbits
presumably derive from the same protostellar disk. However, some
theories of planetary migration---proposed to explain how giant
planets attain short-period orbits---predict occasionally large
misalignments (Chatterjee et al.~2007, Fabrycky \& Tremaine 2007, Wu
et al.~2007, Nagasawa et al.~2008). These theories, as well as the
general history of surprises in this field, provide motivation to
continue measuring exoplanetary spin-orbit alignment.

In this paper, we present a measurement of the RM effect for the
transiting exoplanetary system TrES-2. This system was discovered by
O'Donovan et al.~(2006). It consists of a planet with a mass of
$1.2$~$M_{\rm Jup}$ and radius $1.2$~$R_{\rm Jup}$ orbiting a G0V star
with a period of 2.5~d (O'Donovan et al.~2006, Holman et al.~2007,
Sozzetti et al.~2007). It did not stand out as a promising RM target
because the star is relatively faint ($V=11.4$) and is a slow rotator
($v\sin i_\star = 2.0\pm 1.5$~km~s$^{-1}$; O'Donovan et al.~2006). On
the other hand, the transit occurs at a high impact parameter across
the stellar disk ($b=0.8540\pm 0.0062$; Holman et al.~2007), a
favorable circumstance for this type of measurement (Gaudi \& Winn
2007). Furthermore, in our continuing effort to measure the spin-orbit
angles for a statistically meaningful number of systems, we do not
want to ignore stars with small sky-projected rotation rates. This is
because a small value of $v\sin i_\star$ might be caused by a small
value of $\sin i_\star$, i.e., there might be a large spin-orbit
misalignment. For these reasons, we pursued TrES-2. We describe the
new data in \S~2, the model that we used to interpret the data in
\S~3, and the results in \S~4.

\section{Observations and Data Reduction}

We observed a transit of TrES-2 on UT~2007~April~26 with the Keck~I
10m telescope and the High Resolution Echelle Spectrometer (HIRES;
Vogt et al.~1994). We set up the instrument in the same manner that
has been used consistently for the California-Carnegie planet search
(Butler et al.~1996, 2006). In particular we employed the red
cross-disperser and used the I$_2$ absorption cell to calibrate the
instrumental response and the wavelength scale. The slit width was
$0\farcs85$ and the typical exposure time was 3-4~min, giving a
resolution of about 70,000 and a signal-to-noise ratio (SNR) of
approximately 200~pixel$^{-1}$. We observed the star for 4~hr
bracketing the predicted transit midpoint and obtained a total of 56
spectra, of which 30 were taken during the transit.

We also obtained two iodine-free spectra, with a higher SNR and higher
resolution. We used the sum of these spectra as a template for the
Doppler analysis, which was performed with the algorithm of Butler et
al.~(1996). We estimated the measurement error in the Doppler shift
derived from a given spectrum based on the scatter among the solutions
for individual 2~\AA~sections of the spectrum. The typical error was
6~m~s$^{-1}$. The data are given in Table~1 and plotted in Figs.~1 and
2. Also shown in those figures are data obtained previously by
O'Donovan et al.~(2006), consisting of 11 velocities measured with
Keck/HIRES using a different setup\footnote{Table~3 of O'Donovan et
  al.~(2006) gives incorrect values for the heliocentric Julian dates
  of the velocity measurements. The corrected dates were provided to
  us by D.~Charbonneau (private communication, 2007).}, as well as the
photometric data of Holman et al.~(2007).

\begin{figure}[p]
\epsscale{1.0}
\plotone{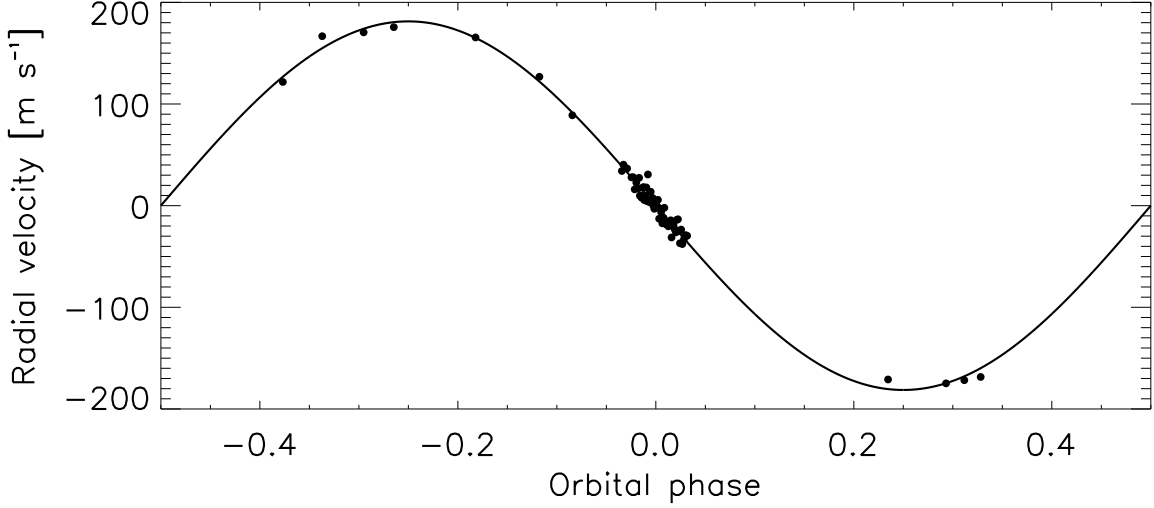}
\caption{
Radial velocity measurements of TrES-2, from this work and from
O'Donovan et al.~(2006), as a function of orbital phase.
The best-fitting values of the systemic velocity have been
subtracted. The solid line is the best-fitting model.
\label{fig:1}}
\end{figure}

\begin{figure}[p]
\epsscale{0.75}
\plotone{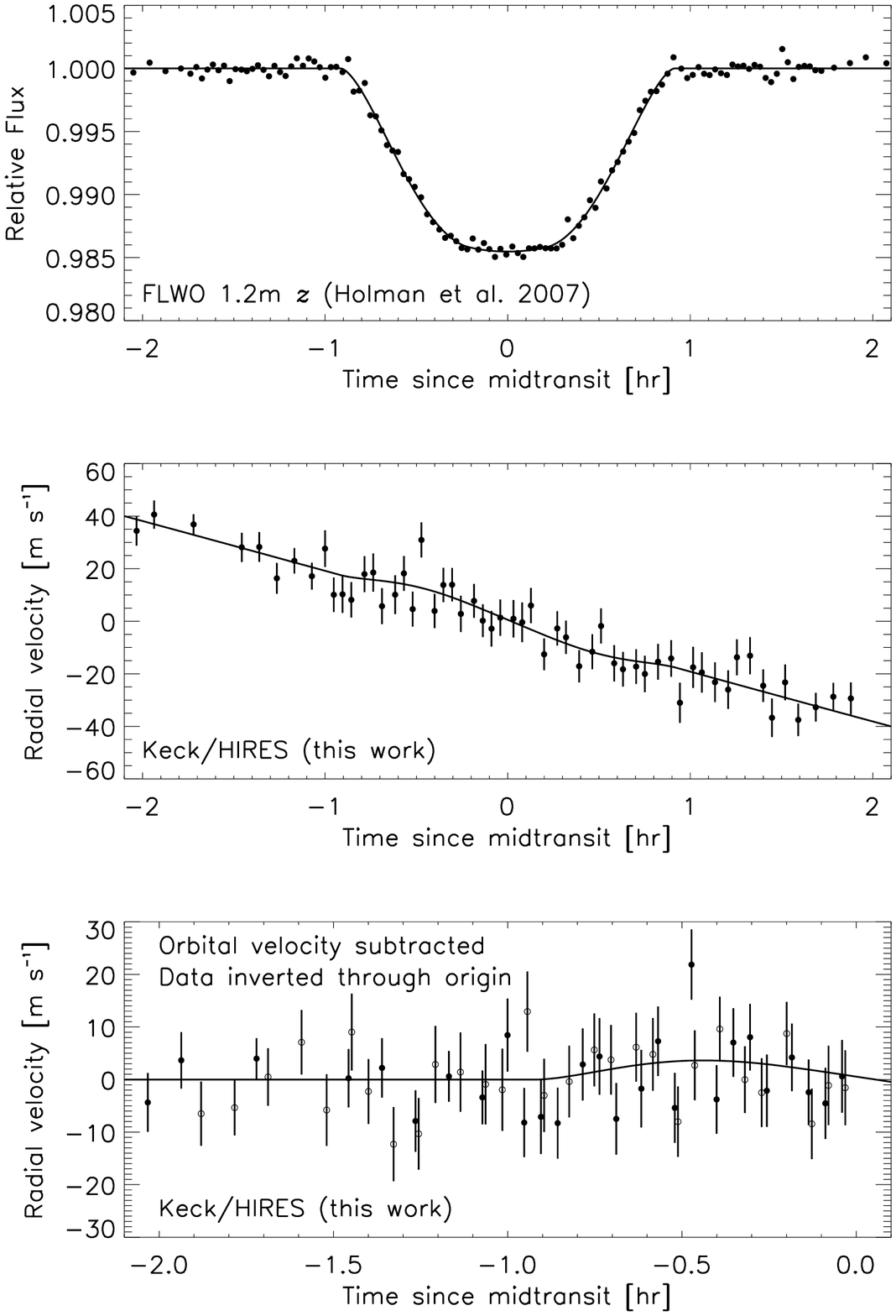}
\caption{ 
{\it Top.} The $z$-band photometry of Holman et al.~(2007), averaged
into 1.5~min bins. The solid line is the best-fitting model. {\it
  Middle.} A close-up of the radial velocity data shown in Fig.~1,
centered on the midtransit time.  {\it Bottom.} Same, but the orbital
velocity has been subtracted and the post-midtransit data ($t > 0$)
have been inverted about the origin ($t\rightarrow -t$ and $\Delta v
\rightarrow -\Delta v$), highlighting the Rossiter-McLaughlin anomaly.
Filled symbols denote data from before midtransit, and open symbols
denote data from after midtransit.
\label{fig:2}}
\end{figure}

\section{The Model}

To determine the projected spin-orbit angle and its uncertainty, we
simultaneously fitted a parametric model to the radial-velocity data
as well as the photometric data of Holman et al.~(2007). We included
the photometric data as a convenient way to account for the
uncertainties in the photometric parameters and their covariances with
the spin-orbit parameters, although in practice the photometric
uncertainties were irrelevant for this system.

The model is based on a circular orbit of a star and planet. The
photometric transit model was identical to the model used by Holman et
al.~(2007). To calculate the anomalous Doppler shift as a function of
the positions of the planet and star, we used the technique of Winn et
al.~(2005): we simulated in-transit spectra, and determined the
Doppler shifts using the same algorithm used on the actual data. The
simulations rely on a template spectrum (described below) that is
meant to mimic the emergent spectrum from a small portion of the
photosphere. At a given moment of the transit, we denote by $\epsilon$
the fractional loss of stellar flux, and we denote by $v_p$ the
line-of-sight velocity of the occulted portion of the stellar disk. To
represent the occulted portion of the stellar spectrum, we scaled the
template spectrum in flux by $\epsilon$ and shifted it in velocity by
$v_p$. We subtracted the scaled and shifted spectrum from a
rotationally-broadened template spectrum and then ``measured'' the
anomalous Doppler shift $\Delta v$. This was repeated for a grid of
$\{\epsilon, v_p\}$, and a polynomial function was fitted to the
resulting grid. We used this polynomial to calculate the anomalous
Doppler shift $\Delta v$ as a function of $\epsilon$ and $v_p$, which
are themselves functions of time. Differential rotation was ignored,
as its effects are expected to be negligible (Gaudi \& Winn 2007).

The template spectrum should be similar to that of TrES-2 but with
slightly narrower lines because of the lack of rotational
broadening. We experimented with two different empirical templates
based on observations of similar stars,\footnote{The two stars were
  HD~38858 ($T_{\rm eff} = 5726$~K, $\log g = 4.51\pm 0.08$,
  [Fe/H]~=~$-0.23\pm 0.04$, $v\sin i_\star = 0.3\pm 0.5$~km~s$^{-1}$)
  and HD~66428 ($T_{\rm eff} = 5752$~K, $\log g = 4.49\pm 0.08$,
  [Fe/H]~=~$+0.31\pm 0.04$, $v\sin i_\star = 0.0\pm
  0.5$~km~s$^{-1}$). The stellar parameters are from the SPOCS catalog
  (Valenti \& Fischer 2005).} finding that both templates gave results
consistent with the function $\Delta v = -\epsilon~v_p$. This function
is consistent with the analytic expressions of Ohta et al.~(2006) and
Gimenez~(2006), even though those analytic expressions do not attempt
to account for the spectral deconvolution. It is simpler than the
quadratic or cubic functions that we have derived for other systems
(Winn et al.~2005, 2006, 2007a). We do not know the reason for the
difference but it is possibly related to the much slower projected
rotation speed of TrES-2.

The fitting statistic was
\begin{eqnarray}
\chi^2 & = &
\sum_{j=1}^{1033}
\left[
\frac{f_j({\mathrm{obs}}) - f_j({\mathrm{calc}})}{\sigma_{f,j}}
\right]^2
+
\sum_{j=1}^{67}
\left[
\frac{v_j({\mathrm{obs}}) - v_j({\mathrm{calc}})}{\sigma_{v,j}}
\right]^2
,
\end{eqnarray}
where $f_j$(obs) and $\sigma_{f,j}$ are the flux measurements and
uncertainties of Holman et al.~(2007), and $v_j$(obs) and
$\sigma_{v,j}$ are the radial-velocity measurements and uncertainties
from our new data and from O'Donovan et al.~(2006). The two model
parameters relating to the RM effect are the line-of-sight stellar
rotation velocity ($v \sin i_\star$), and the angle between the
projected stellar spin axis and orbit normal ($\lambda$). The
projected spin-orbit angle $\lambda$ ranges from $-180\arcdeg$ to
$+180\arcdeg$, and is measured counterclockwise on the sky from the
projected stellar rotational angular-momentum vector to the projected
orbital angular-momentum vector (see Ohta et al.~2007 or Gaudi \& Winn
2007 for a diagram). If we define stellar ``north'' by the sky
projection of the stellar angular-momentum vector, then when
$\lambda=0\arcdeg$ the axes are aligned and the planet moves directly
``eastward'' across the face of the star, for $0\arcdeg < \lambda <
90\arcdeg$ the planet moves ``northeast,'' and so forth.

The other model parameters were the planetary mass ($M_p$); the
stellar and planetary radii ($R_\star$ and $R_p$); the orbital
inclination ($i$); the mid-transit time ($T_c$); and an additive
constant velocity for each of the two different velocity data sets
($\gamma_1$ and $\gamma_2$). We allowed our velocities to have a
different additive constant from the velocities of O'Donovan et
al.~(2006) in order to account for systematic differences in the
spectrograph setup and reduction procedures. We fixed the orbital
period to be $2.47063$~days (Holman et al.~2007). We used a Markov
Chain Monte Carlo algorithm to solve for the model parameters and
their confidence limits, with uniform priors on all parameters. This
algorithm and our implementation of it are described in detail
elsewhere (see, e.g., Winn et al.~2007b). The minimum $\chi^2$ is
1127.6, with 1091 degrees of freedom, giving $\chi^2/N_{\rm dof} =
1.034$ and indicating an acceptable fit.

\section{Results}

The RM effect is certainly not obvious in Fig.~1, which shows the
entire spectroscopic orbit. It is not even very obvious in the middle
panel of Fig.~2, which focuses on the velocity data around the time of
transit. However, our analysis shows that the RM effect was indeed
detected. As mentioned above, for the best-fitting model, $\chi^2_{\rm
  min} = 1127.6$. If the parameter $v\sin i_\star$ is set equal to
zero, thereby neglecting the RM effect, then $\chi^2_{\rm min} =
1135.8$, with the increase of $\Delta\chi^2 = 8.2$ arising from the
velocity data during the transit. We conclude that the RM effect was
detected with a signal-to-noise ratio (SNR) of approximately
$\sqrt{8.2} = 2.9$. Gaudi \& Winn (2007) have given analytic formulas
for the signal-to-noise ratio of RM observations as a function of the
system and telescope parameters, under the assumption of Gaussian
velocity errors. Using their Eqn.~(26) for this case, the forecasted
SNR is 2.9, in agreement with the actual SNR.

One might wonder how much this result was influenced by the inclusion
of the photometric data. To check on this, we tried setting aside the
photometric data and fitting only the 67 radial-velocity data
points. We fixed the photometric parameters ($M_p$, $R_p$, $R_\star$,
$i$, $T_c$, and $P$) at the values determined previously. In this case
we found $\chi^2_{\rm min} = 63.7$. If $v\sin i_\star$ is set equal to
zero, then $\chi^2_{\rm min} = 71.9$, giving $\Delta\chi^2 = 8.2$,
just as in the full model fit. This confirms that the lowered $\chi^2$
is an effect of a better fit to the transit velocities, and that the
uncertainties in the photometric parameters are negligible in this
instance.

The best-fitting model parameters are also consistent with good
alignment of the spin and the orbit. Specifically, we find $\lambda =
-9\pm 12$~deg, and $v\sin i_\star = 1.0\pm 0.6$~km~s$^{-1}$, where the
quoted values are the medians of the {\it a posteriori}\,
distributions returned by the MCMC algorithm, and the error bars
represent 68\% confidence limits. Table~2 gives these results, along
with some other revelant system parameters of TrES-2, for
convenience. Visually, the RM effect is more apparent in the bottom
panel of Fig.~2, in which the orbital velocity has been subtracted
from the data, and the sampling rate has been effectively doubled by
inverting the data through the origin ($t \rightarrow -t$ and $\Delta
v \rightarrow -\Delta v$). This works because for $\lambda\approx 0$,
the RM waveform is antisymmetric about the origin.

Figure~3 shows the {\it a posteriori}\, probability distribution for
$\lambda$ and the joint distribution of $\lambda$ and $v\sin
i_\star$. The distribution for $\lambda$ resembles a slightly
asymmetric Gaussian function to which is added a low-level uniform
probability distribution.  Although only the region from $-90\arcdeg$
to $+90\arcdeg$ is shown in Fig.~3, this low-level uniform
distribution extends all the way from $-180\arcdeg$ to
$+180\arcdeg$. The uniform background corresponds to the very lowest
allowed values of $v\sin i_\star$. This makes sense because when the
rotation rate is zero, the Rossiter anomaly vanishes and $\lambda$ is
irrelevant. Values of $\lambda$ between $-90\arcdeg$ and $+90\arcdeg$
correspond to prograde orbits, for which the stellar and orbital
angular momenta are in the same half-plane. The integrated probability
between $-90\arcdeg$ and $+90\arcdeg$ is 98\%. We conclude that the
TrES-2 orbit is prograde with 98\% confidence. As an illustration of
the constraints provided by our analysis, Fig.~4 shows a drawing of
the face of the star and the orbit of the transiting planet.

Our result for $v\sin i_\star$ is in agreement with the value reported
by O'Donovan et al.~(2006), $2.0\pm 1.5$~km~s$^{-1}$, which was based
on an analysis of the line broadening in an out-of-transit spectrum.
This finding is also supported by an analysis of our own
out-of-transit, iodine-free spectra, using the {\it Spectroscopy Made
  Easy}\, (SME) software package of Valenti \& Piskunov~(1996). The
automated analysis gave a formal result of $v\sin i_\star = 0.5\pm
0.5$~km~s$^{-1}$, although the true uncertainty may be larger, since
with a disk-integrated spectrum of such a slow rotator it is difficult
to disentangle the effects of rotation, macroturbulence,
microturbulence, and the instrumental profile. In particular, the SME
code assumes ``typical'' values for the turbulent broadening
mechanisms that are of the same magnitude as the rotation speed of
TrES-2 (see \S~4.2-4.4 of Valenti \& Fischer 2005).\footnote{We
  investigated the consequences of accepting the SME result at face
  value, by imposing a Gaussian prior constraint on the $v\sin
  i_\star$ parameter with mean 0.5~km~s$^{-1}$ and standard error
  0.5~km~s$^{-1}$. In that case, the MCMC analysis gave
  68\%-confidence ranges of $-31$ to $1$~deg for $\lambda$ and $0.3$
  to $1.1$~km~s$^{-1}$ for $v\sin i_\star$, and showed that the orbit
  is prograde with 95\% confidence. The constraint on $\lambda$ was
  weakened because the SME result favors slower rotation rates, for
  which the sensitivity of the RM waveform to $\lambda$ is reduced.}

\begin{figure}[p]
\epsscale{0.75}
\plotone{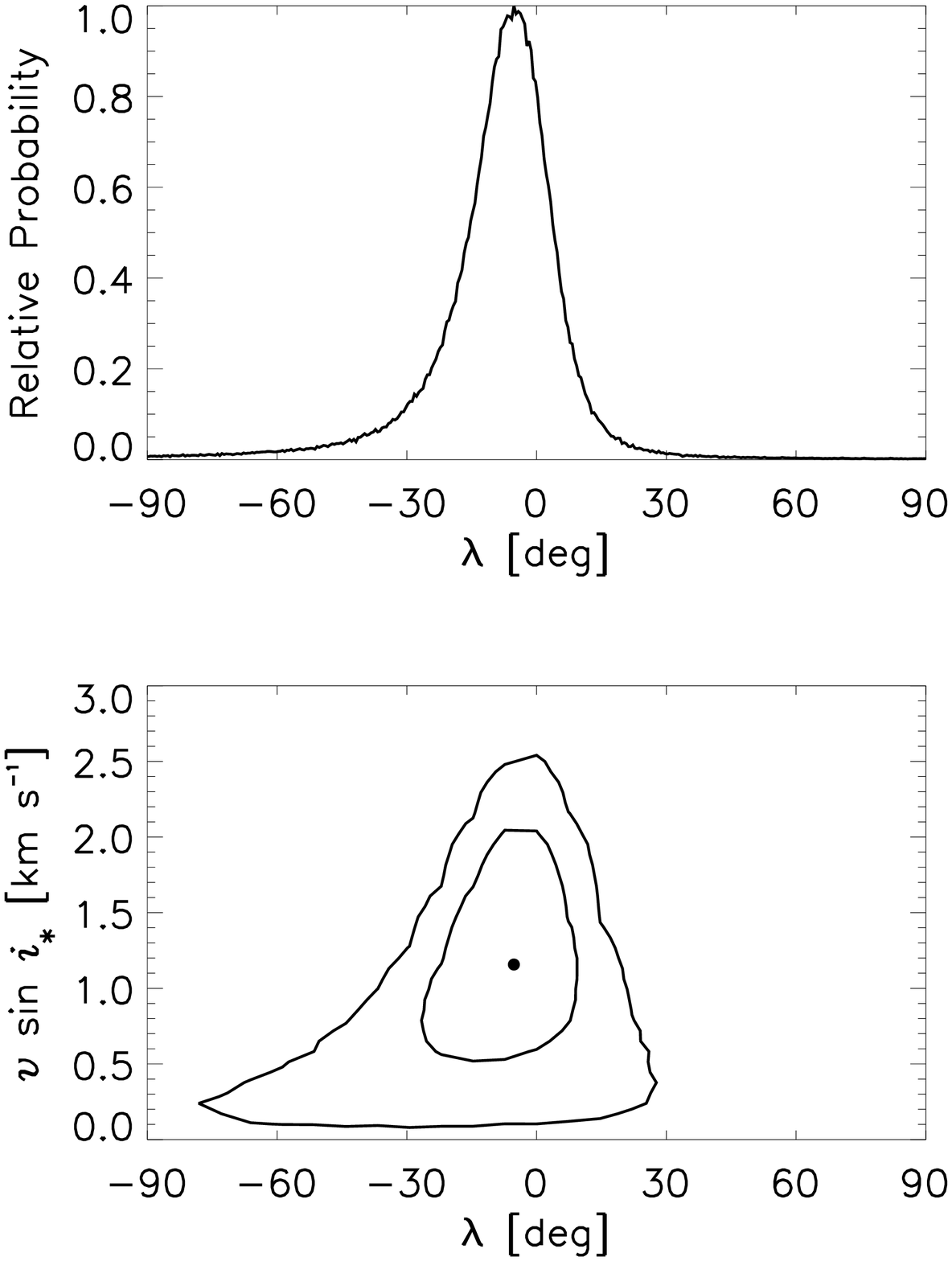}
\caption{
{\it Top.}---The probability distribution for $\lambda$, the angle
between the sky projections of the orbital axis and the stellar
rotation axis. {\it Bottom.}---The joint probability distribution
of $\lambda$ and $v\sin i_\star$. The solid dot shows the best-fitting
values. The contours represent 68\% and 95\% confidence limits.
\label{fig:3}}
\end{figure}

\begin{figure}[p]
\epsscale{0.75}
\plotone{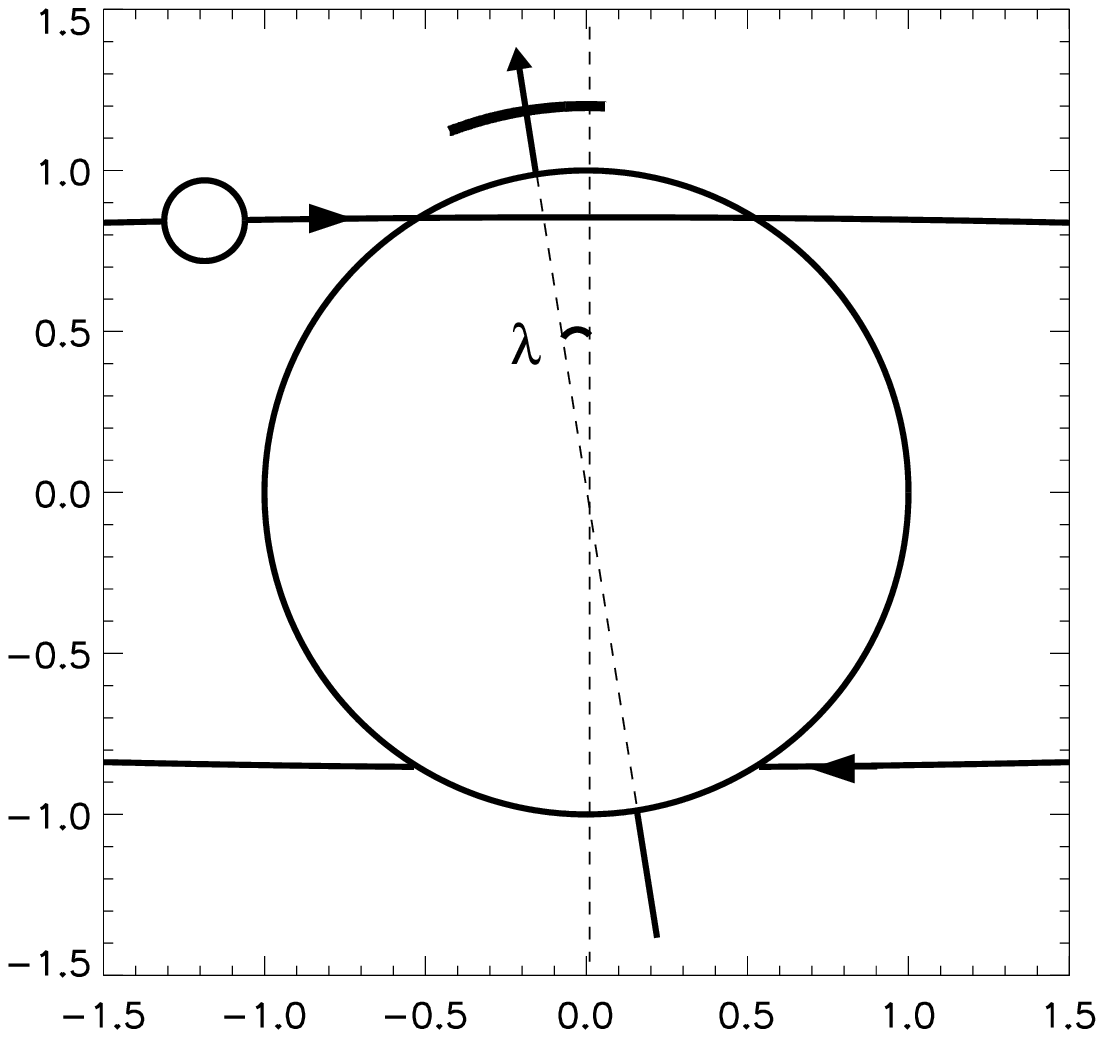}
\caption{
Scale drawing of the TrES-2 system.  The relative radii of the bodies
and the impact parameter of the transit are taken from our
best-fitting model.  The ``north pole'' of the star is drawn with an
arrow, and the curved arc shows the 68\%-confidence region for its
orientation.  The angle $\lambda$ is measured clockwise from the
projected orbit normal vector (vertical dashed line) to the projected
stellar north pole (tilted dashed line). The best-fitting value of
$\lambda$ is negative.
\label{fig:4}}
\end{figure}

\section{Summary and Discussion}

We have monitored the apparent Doppler shift of TrES-2 throughout a
transit of its giant planet and we have detected the
Rossiter-McLaughlin effect. Using the available photometric and
spectroscopic data, we have found good evidence that the orbit is
prograde, as are the other 6 systems that have been measured (with the
possible exception of HD~17156), and as are the planets in the Solar
system. In this sense, our results for TrES-2 are not surprising.
However, as mentioned in \S~1, some theories of planet migration do
predict occasionally large misalignments. For example, Nagasawa et
al.~(2008) investigated a scenario in which a planet is scattered into
an eccentric, inclined orbit with a small periastron distance (as
envisioned earlier by Rasio \& Ford 1996 and Marzari \&
Weidenschilling 2002), and subsequently a more distant planet forces
Kozai oscillations in the inner planet's eccentricity and
inclination. If the periastron distance is small enough during the
high-eccentricity phases, the orbit may circularize at a small orbital
distance with a substantial inclination. Nagasawa et al.~(2008) found
that this migration mechanism produces a very broad range of final
inclinations, including a significant fraction of retrograde
orbits. Of course, prograde orbits are also permitted in this
scenario, and our finding of a prograde orbit for TrES-2 cannot be
taken as evidence against this mechanism. We raise the issue only to
show that a prograde orbit was not a foregone conclusion.

Furthermore, we have shown it is possible to glean this information
and measure the projected spin-orbit angle to within $12\arcdeg$, even
for an 11th magnitude star with a slow projected rotation rate. A
potentially important application of the RM effect is the detection of
planets that are too small to be readily detected using other types of
ground-based data. For example, in many cases of terrestrial planets
detected by the {\it Corot}\, or {\it Kepler}\, satellites, it will be
easier to observe the RM effect than to observe the star's orbital
Doppler shift (and thereby measure the planet's mass). The theory
underlying this idea has been discussed by Welsh et al.~(2004) and
Gaudi \& Winn~(2007).

The present work serves to illustrate this point with actual data. If
TrES-2 had a rotation rate of 5~km~s$^{-1}$ instead of 1~km~s$^{-1}$,
but all other stellar and orbital parameters were the same, then the
quantity and quality of data presented in this paper would permit a
$\sim$3$\sigma$ detection of a planet with a radius $\sim$$\sqrt{5}$
times smaller than TrES-2, or $\sim$6 Earth radii. If the transit were
equatorial instead of grazing (the best configuration for detecting
the effect, although not for assessing spin-orbit alignment), the
duration of the transit would be longer by a factor of $\sim$2 and the
amplitude of the RM effect would be larger by a factor of $\sim$2,
leading to another factor-of-2 improvement in the detectable planet
radius ($\sim$3~$R_\oplus$). Such a planet would produce a photometric
transit depth of only $8\times 10^{-4}$, which is smaller than the
transit depth of any known transiting planet.

\acknowledgments We thank G.~Marcy for advice and encouragement, and
D.~Charbonneau for helpful conversations. We are grateful for support
from the NASA Keck PI Data Analysis Fund (JPL 1326712). We recognize
and acknowledge the very significant cultural role and reverence that
the summit of Mauna Kea has always had within the indigenous Hawaiian
community. We are most fortunate to have the opportunity to conduct
observations from this mountain. Access to the Keck telescopes for
this project was through the Telescope System Instrumentation Program,
and was supported by AURA through the National Science Foundation
under AURA Cooperative Agreement AST 0132798 as amended.

\begin{deluxetable}{lcccc}
\tabletypesize{\tiny}
\tablecaption{Radial Velocities of TrES-2\label{tbl:rv}}
\tablewidth{0pt}

\tablehead{
\colhead{HJD} & \colhead{Radial Velocity [m~s$^{-1}$]} & \colhead{Measurement Uncertainty [m~s$^{-1}$]}
}

\startdata
 $  2454216.96599$ & $          42.59$ & $           5.86$ \\
 $  2454216.96998$ & $          38.52$ & $           5.76$ \\
 $  2454216.97930$ & $          37.06$ & $           5.43$ \\
 $  2454216.98973$ & $          26.69$ & $           5.76$ \\
 $  2454216.99368$ & $          27.35$ & $           5.82$ \\
 $  2454216.99769$ & $          21.46$ & $           5.87$ \\
 $  2454217.00168$ & $          22.74$ & $           5.69$ \\
 $  2454217.00564$ & $          17.82$ & $           5.75$ \\
 $  2454217.00876$ & $          31.44$ & $           6.16$ \\
 $  2454217.01102$ & $          16.32$ & $           6.06$ \\
 $  2454217.01327$ & $          12.64$ & $           6.24$ \\
 $  2454217.01552$ & $           6.15$ & $           6.05$ \\
 $  2454217.01779$ & $          17.40$ & $           6.13$ \\
 $  2454217.02003$ & $          19.25$ & $           6.35$ \\
 $  2454217.02229$ & $           5.79$ & $           6.24$ \\
 $  2454217.02453$ & $          14.41$ & $           6.35$ \\
 $  2454217.02681$ & $          23.50$ & $           6.26$ \\
 $  2454217.02905$ & $           6.40$ & $           6.19$ \\
 $  2454217.03131$ & $          35.58$ & $           6.27$ \\
 $  2454217.03356$ & $           6.90$ & $           5.92$ \\
 $  2454217.03580$ & $          14.89$ & $           6.07$ \\
 $  2454217.03803$ & $          16.89$ & $           6.09$ \\
 $  2454217.04040$ & $           4.04$ & $           6.25$ \\
 $  2454217.04266$ & $           7.37$ & $           6.17$ \\
 $  2454217.04492$ & $           3.26$ & $           5.91$ \\
 $  2454217.04715$ & $           0.45$ & $           6.44$ \\
 $  2454217.04940$ & $           3.51$ & $           6.27$ \\
 $  2454217.05165$ & $          -0.38$ & $           6.31$ \\
 $  2454217.05403$ & $           2.49$ & $           6.37$ \\
 $  2454217.05648$ & $           7.42$ & $           6.13$ \\
 $  2454217.05907$ & $          -8.45$ & $           6.16$ \\
 $  2454217.06167$ & $          -4.76$ & $           6.17$ \\
 $  2454217.06425$ & $          -7.03$ & $           6.06$ \\
 $  2454217.06684$ & $         -11.21$ & $           6.06$ \\
 $  2454217.06957$ & $          -8.69$ & $           6.32$ \\
 $  2454217.07214$ & $          -6.58$ & $           6.15$ \\
 $  2454217.07473$ & $         -17.63$ & $           6.14$ \\
 $  2454217.07730$ & $         -19.40$ & $           6.14$ \\
 $  2454217.07991$ & $         -25.50$ & $           6.18$ \\
 $  2454217.08250$ & $         -16.69$ & $           6.14$ \\
 $  2454217.08513$ & $         -11.51$ & $           6.33$ \\
 $  2454217.08767$ & $         -16.22$ & $           6.21$ \\
 $  2454217.09031$ & $         -28.52$ & $           6.52$ \\
 $  2454217.09287$ & $         -18.49$ & $           6.62$ \\
 $  2454217.09545$ & $          -9.36$ & $           6.50$ \\
 $  2454217.09806$ & $         -25.95$ & $           6.43$ \\
 $  2454217.10065$ & $         -25.17$ & $           6.26$ \\
 $  2454217.10335$ & $         -15.71$ & $           6.36$ \\
 $  2454217.10608$ & $         -15.20$ & $           6.34$ \\
 $  2454217.10868$ & $         -21.00$ & $           6.26$ \\
 $  2454217.11124$ & $         -30.07$ & $           6.31$ \\
 $  2454217.11386$ & $         -22.90$ & $           6.24$ \\
 $  2454217.11709$ & $         -35.76$ & $           5.99$ \\
 $  2454217.12110$ & $         -33.87$ & $           5.88$ \\
 $  2454217.12509$ & $         -29.70$ & $           5.79$ \\
 $  2454217.12911$ & $         -26.81$ & $           5.88$
\enddata 

\tablecomments{Column 1 gives the Heliocentric Julian Date at the
  photon-weighted midexposure time, i.e., weighted by the photon count
  rate recorded by the HIRES exposure meter.}

\end{deluxetable}

\begin{deluxetable}{lccc}
\tabletypesize{\normalsize}
\tablecaption{System Parameters of TrES-2\label{tbl:params}}
\tablewidth{0pt}

\tablehead{
\colhead{Parameter} & \colhead{Value} & \colhead{68\% Confidence Limits} & \colhead{References}
}
\startdata 
$P$~[d]                & $2.470621$ & $\pm 0.000017$ & 1  \\
$T_c$ [HJD]            & $2453957.63479$ & $\pm 0.00038$ & 1  \\
$(R_p/R_\star)^2$          & $0.0157$ & $\pm 0.0003$             & 1 \\
$b \equiv a\cos i/R_\star$ & $0.8540$ & $\pm 0.0062$             & 1 \\
$M_\star~[M_\odot]$        & $0.980$  & $\pm 0.062$          & 2 \\
$R_\star~[R_\odot]$        & $1.000$  & $+0.036$, $-0.033$ & 2 \\
$M_p~[M_{\rm Jup}]$        & $1.198$  & $\pm 0.053$          & 2 \\
$R_p~[R_{\rm Jup}]$        & $1.220$  & $+0.045$, $-0.042$ & 2 \\
$v\sin i_\star$~[km~s$^{-1}$] & $1.0$    & $\pm 0.6$                & This work \\
$\lambda$~[deg]            & $-9$   & $\pm 12$   & This work
\enddata
\tablecomments{(1) Holman et al.~(2007); (2) Sozzetti et al.~(2007).}

\end{deluxetable}

\end{document}